# BIDIRECTIONALLY COUPLED VARIABLE MULTIPLE TIME DELAY SYSTEMS AND INVERSE CHAOS SYNCHRONIZATION

E.M.Shahverdiev

Institute of Physics, H.Javid Avenue,33, Baku, AZ1143, Azerbaijan

e-mail:shahverdiev@physics.ab.az

## ABSTRACT

We present the first report on inverse chaos synchronization between bi-directionally non-linearly and linearly coupled variable multiple time delay Ikeda systems. The results are of certain importance in secure chaos-based communication systems.

## 1. INTRODUCTION

In recent years the presence of chaotic vibrations and its control in nonlinear dynamical systems in physics, power electronics, ecology, economics and so on, has been extensively demonstrated and is now well established, see, e.g. references in [1-2]. Chaos synchronization [3] is another basic feature in nonlinear science and is one of fundamental importance in a variety of complex physical, chemical, and biological systems, see e.g. references in [1-2]. Equivalence of synchronization and control of chaotic systems was emphasized in [2]. Potential application areas of chaos synchronization include secure communications, optimization of nonlinear system performance, information processing, and pattern recognition phenomena [1-2].

Recently, delay differential equations [4] have attracted much attention in the field of nonlinear dynamics. The high complexity of the multiple time-delayed systems can provide a new architecture for enhancing message security in chaos based encryption systems [5]. In such communication systems message decoding would require chaos synchronization between multiple time-delayed transmitter and receiver systems [1-3]. Variable multiple time delay systems are further generalization of the fixed time delay systems [6]. Such variations of time delays could be intentionally or as a result of fluctuations. In a word, modulated time delay systems could be more realistic models of interacting systems. Investigation of synchronization possibilities in such systems are of certain importance.

There are different types of synchronization in interacting chaotic systems [1-2]. Complete, generalized, phase, lag and anticipating synchronizations of chaotic vibrations have been described theoretically and observed experimentally. Complete synchronization implies coincidence of states of interacting systems, $y(t)=x(t)$ [3]; a generalized synchronization is defined as the presence of some functional relation between the states of response and drive, i.e. $y(t)=F(x(t))$; phase synchronization means entrainment of phases of chaotic oscillators, $n\Phi_x - m\Phi_y =$ const, (n and m are integers) whereas their amplitudes remain chaotic and uncorrelated; lag synchronization appears as a coincidence of shifted-in-time states of two systems, $y(t)=x(t-\tau)$ with positive $\tau$; anticipating synchronization also appears as a coincidence of shifted-in-time states of two coupled systems, but in this case, in contrast to lag synchronization, the driven system anticipates the driver, $y(t)=x(t+\tau)$, $\tau>0$ [1-2]. An experimental observation of anticipating synchronization in external cavity laser diodes has been reported in [7]. For inverse synchronization [8], a time-delayed chaotic system x is coupled with an another system y in such a way that one system's dynamics synchronize to the inverse state of the other system or vice-versa: $x(t)=-y(t)$.

Most chaos based communication techniques use synchronization in unidirectional master-slave system. Such a coupling scheme prevents the messages being exchanged between the sender and receiver. A two way transmission of signals requires bidirectional coupling. With this in mind this paper presents the first report of the inverse chaos synchronization between bi-directionally non-linearly and linearly coupled modulated multiple time delayed Ikeda models with two feedbacks.

II. SYSTEM MODEL

First consider inverse synchronization between bi-directionally nonlinearly coupled variable time-delayed Ikeda systems with two feedbacks,

$$\frac{dx}{dt} = -\alpha x - m_1 \sin x(t-\tau_1) - m_2 \sin x(t-\tau_2) + K_y \sin y(t-\tau_3) \quad (1)$$

$$\frac{dy}{dt} = -\alpha y - m_3 \sin y(t-\tau_1) - m_4 \sin y(t-\tau_2) + K_x \sin x(t-\tau_3) \quad (2)$$

This investigation is of considerable practical importance, as the equations of the class B lasers with feedback (typical representatives of class B are solid-state, semiconductor, and low pressure $CO_2$ lasers) can be reduced to an equation of the Ikeda type, see e.g. references in [8]. The Ikeda model was introduced to describe the dynamics of an optical bi-stable resonator, playing an important role in electronics and physiological studies and is well-known for delay-induced chaotic behavior, see e.g. references in [8].

Physically x is the phase lag of the electric field across the resonator; $\alpha$ is the relaxation coefficient for the driving x and driven y dynamical variables; $\tau_{1,2} = \tau_{01,02} + x_1(t) \tau_{a1,a2} \sin(\varpi_{1,2} t)$ are the variable feedback loop delay times; $\tau_3 = \tau_{03} + x_1(t) \tau_{a3} \sin(\varpi_3 t)$ is the variable time of flight between systems x and y; $\tau_{01,02,03}$ are the zero-frequency component, $\tau_{a1,a2,a3}$ are the amplitude, $\frac{\varpi_{1,2,3}}{2\pi}$ are the frequency of the modulations; $x_1(t)$ is the output of system (1) ($K_y = 0$) for constant time delays, i.e. $\tau_1 = \tau_{01}$, $\tau_2 = \tau_{02}$; $m_{1,2}$ and $m_{3,4}$ are the feedback strengths for x and y systems, respectively; $K_{x,y}$ are the coupling strengths between the systems. Variable time delays $\tau_{1,2,3}$ are chosen in such a way to include both chaotic and non-chaotic components.

Before considering the case of modulated time delays, we present the conditions for inverse synchronization for fixed time delays, i.e. $\varpi_{1,2,3}=0$. The case for fixed time delays was investigated in [9] and here we briefly reproduce the results from [9]. The key difference between this study and ref. [9] is that here we study the case of inverse chaos synchronization between variable multiple time delays systems, whereas the authors of [9] dealt with the inverse chaos synchronization between the systems with fixed time delays. As mentioned above, in real situations modulated time delays systems could be more realistic models of the communicating systems.

As established in [9] systems (1) and (2) can be synchronized on the inverse synchronization regime

$$x=-y \tag{3}$$

under the existence conditions

$$m_1=m_3, m_2=m_4, K_x = K_y. \tag{4}$$

The sufficient stability condition for the synchronization regime $x=-y$ can be investigated by the use of Razumikhin-Lyapunov functional approach and is found to be of the form [9]:

$$\alpha > |m_3| + |m_4| + |K_y| \tag{5}$$

At first glance, the sufficient stability condition (5) for chaos synchronization is difficult to satisfy, as higher values of $\alpha$ could render the dynamics trivial, i.e. non-chaotic. Fortunately there is a way out of this impasse. The key point is that the stability conditions derived from the Lyapunov-Razumikhin approach is a sufficient one: it assures a high quality synchronization for a coupling strength estimated from the stability condition, but does not forbid the possibility of synchronization with smaller coupling strengths. Below we demonstrate numerically that one can still achieve high quality chaos synchronization in mutually coupled Ikeda systems with double time delays without the condition (5) being fulfilled.

In the case of variable time delays establishing the stability conditions for the synchronization is not as straightforward as for the constant time delays. Having in mind that for $\varpi_{1,2,3}=0$ we obtain a case of constant time delays, then as an initial guess one can benefit from the existence conditions for the constant time delays case. It is our conjecture that high quality synchronization x=-y could be obtained if the parameters satisfy conditions (4). As evidenced by the numerical simulations below, this conjecture is found to be well-based.

Now consider linearly bi-directionally coupled Ikeda systems:

$$\frac{dx}{dt} = -\alpha x - m_1 \sin x(t-\tau_1) - m_2 \sin x(t-\tau_2) + Ky \qquad (6)$$

$$\frac{dy}{dt} = -\alpha y - m_3 \sin y(t-\tau_1) - m_4 \sin y(t-\tau_2) + Kx \qquad (7)$$

For this kind of coupling the existence conditions for inverse synchronization regime (3) are: $m_1=m_3$ and $m_2=m_4$ with sufficient stability condition: $\alpha > |m_3| + |m_4| + K$.

III. NUMERICAL SIMULATIONS

In numerical simulations to characterize the quality of synchronization we calculate the cross-correlation coefficient C [10]

$$C = <(x(t)-<x>)(y(t+\Delta t)-<y>)>(<(x(t)-<x>)^2><(y(t+\Delta t)-<y>)^2>)^{1/2} \qquad (8)$$

where x and y are the outputs of the interacting laser systems; the brackets <.> represent the time average; $\Delta t$ is a time shift between laser outputs. This coefficient indicates the quality of synchronization: C=-1 means perfect inverse synchronization; for x=-y synchronization $\Delta t=0$.

As mentioned above, in chaos based communication schemes synchronization between the transmitter and receiver systems are vital for message decoding. With this in mind in the remainder of the paper we focus on the inverse synchronization between bi-directionally non-linearly and linearly coupled Ikeda systems with double variable time delays.

Figure 1 portrays time series of the Ikeda system x (solid line) and system y (dotted line) for inverse chaos synchronization x=-y between non-linearly coupled systems, equations (1) and (2) for variable feedback time delays $\tau_1(t)=3 + 2x_1(t)\sin(0.15t)$, $\tau_2(t)=5 + 2x_1(t)\sin(0.15t)$ and variable coupling time delay $\tau_3(t)=7 + 2x_1(t)\sin(0.15t)$ with parameter values as $\alpha=3$, $m_1=m_3=3.1$, $m_2=m_4=2.5$, $K_x=K_y=0.03$. $x_1(t)$ is the solution of equation (1) ($K_y=0$) for $\tau_{01}=3$ and $\tau_{02}=5$. Figure 2 depicts synchronization error dynamics $\Delta=x + y$ versus time for parameters as in figure 1. C =0.99 is the cross-correlation coefficient between the transmitter x and receiver y system outputs. It is noted that the parameters used here in the numerical simulations satisfy the existence condition (4), but fail to satisfy the sufficient stability condition (5). Nevertheless high quality synchronization is achieved. Indeed, the value of the cross-correlation coefficient C testify to the high quality chaos synchronization, which is vital for information processing in chaos-based communication systems.

Next we present the results of numerical simulations for the linearly coupled Ikeda systems. Figure 3 presents time series of the Ikeda system x (solid line) and system y (dotted line) for inverse chaos synchronization x=-y between linearly coupled systems, equations (6) and (7) for variable feedback time delays $\tau_1(t)=1 + 0.5x_1(t)\sin(0.1t)$, $\tau_2(t)=3 + 0.5x_1(t)\sin(0.1t)$ with parameter values as $\alpha=4$, $m_1=m_3=3.2$, $m_2=m_4=4.5$, $K=0.01$. $x_1(t)$ is the solution of equation (6) ($K=0$) for $\tau_{01}=1$ and $\tau_{02}=3$. Figure 4 shows receiver output y versus transmitter output x for parameters as in figure 3. C =0.99 is the cross-correlation coefficient between the transmitter x and receiver y system outputs. As in the case of non-linear coupling despite the failure to satisfy the sufficient stability condition $\alpha > |m_3| + |m_4| + K$, as testified by the value of cross-correlation coefficient C=0.99 we were able to achieve high quality chaos synchronization.

IV. CONCLUSIONS

To summarize we have reported on inverse chaos synchronization in bi-directionally non-linearly and linearly coupled variable multiple time delayed Ikeda systems. These results are of certain practical importance in secure chaos-based communication systems.

V. ACKNOWLEDGEMENTS

This research was supported by a Marie Curie Action within the 6$^{th}$ European Community Framework Programme Contract N:MIF2-CT-2007-039927-980065(Reintegration Phase).

Figure Captions

FIG.1. Numerical simulation of bi-directionally non-linearly coupled variable time delay systems, Equations (1-2) for $\alpha = 3$, $m_1=m_3=3.1$, $m_2=m_4=2.5$, $K_x = K_y = 0.03$, and $\tau_1(t)=3 + 2x_1(t)\sin(0.15t)$, $\tau_2(t)=5 + 2x_1(t)\sin(0.15t)$, and $\tau_3(t)=7 + 2x_1(t)\sin(0.15t)$. $x_1(t)$ is the solution of equation (1) ($K_y=0$) for $\tau_{01}=3$ and $\tau_{02}=5$. Inverse synchronization: Time series of x system (solid line) and y system (dotted line). Dimensionless units.

FIG.2. Numerical simulation of bi-directionally coupled variable time delay systems, Equations (1-2). Error dynamics, $\Delta = x + y$ versus time. The parameters are as in figure 1. C is the cross-correlation coefficient between the Ikeda systems. Dimensionless units.

FIG.3. Numerical simulation of bi-directionally linearly coupled variable time delay systems, Eqs.(6-7) for $\alpha = 4$, $m_1=m_3=3.2$, $m_2=m_4=4.5$, $K = 0.01$ and $\tau_1(t)=1 + 0.5x_1(t)\sin(0.1t)$, $\tau_2(t)=3 + 0.5x_1(t)\sin(0.1t)$. $x_1(t)$ is the solution of equation (6) ($K=0$) for $\tau_{01}=1$ and $\tau_{02}=3$. Inverse synchronization: Time series of x system (solid line) and y system (dotted line). Dimensionless units.

FIG.4. Numerical simulation of mutually linearly coupled variable time delay Ikeda systems, equations (6-7). Correlation plot between x and y. C is the correlation coefficient between x and y. Dimensionless units.

REFERENCES

1. Schöll, E., and Schuster, H.G.,(Eds.), 2007, Handbook of Chaos Control,Wiley-VCH, Weinheim, Germany.

2. Boccaletti, S., Kurths, J., Osipov, G., Valladares, D.L. and Zhou C.S.,2002, "The synchronization of chaotic systems," Physics Reports 366,1-100; Konnur, R., 1996,"Equivalence of synchronization and control of chaotic systems," Phys.Rev.Lett.77, 2937-2940.

3. Pecora, L.M., and Carroll, T.L., 1990, "Synchronization in chaotic systems," Phys.Rev.Lett.64, 821-824.

4. Hale, J.K., and Lunel, S.M.V., 1993, Introduction to Functional Differential Equations, Springer, New York, USA.

5. Shahverdiev, E.M., and Shore K.A., 2008, "Chaos synchronization regimes in multiple-time-delay semiconductor lasers," Phys. Rev. E 77, 057201(1-4).

6. Shahverdiev, E.M., and Shore K.A.,2009, "Impact of modulated multiple optical feedback time delays on laser diode synchronization," Opt.Comm.282, 3568-3572.

7. Sivaprakasam, S., Shahverdiev, E.M., Spencer P.S. and Shore K.A.,2001, "Experimental demonstration of anticipating synchronization in chaotic semiconductor lasers with optical feedback," Phys.Rev.Lett.87, 154101(1-4). Masoller, C.,2001, "Anticipating in the synchronization of chaotic semiconductor lasers with optical feedback," Phys. Rev. Lett.86, 2782-2785.

8. Shahverdiev, E.M., Sivaprakasam, S. and Shore K.A., 2002, "Inverse anticipating chaos synchronization," Phys.Rev.E66, 017204(1-4); Njah, A.N.,and Vincent U.E., 2009, "Synchronization and anti-synchronization of chaos in an extended Bonh{\"o}ffer-van der Pol oscillator using active control," J. of Sound and Vibration319, 41-49.

9.Shahverdiev, E.M., Nuriev, R.A., Hashimov, R.H., Hashimova, L.H., Huseynova, E.M.and Shore K.A.,2006, "Inverse chaos synchronization in linearly and nonlinearly coupled systems with multiple time delays," Chaos, Solitons and Fractals29,838-844.

10.Haykin S.,1994, Communication Systems, Wiley, New York,USA.

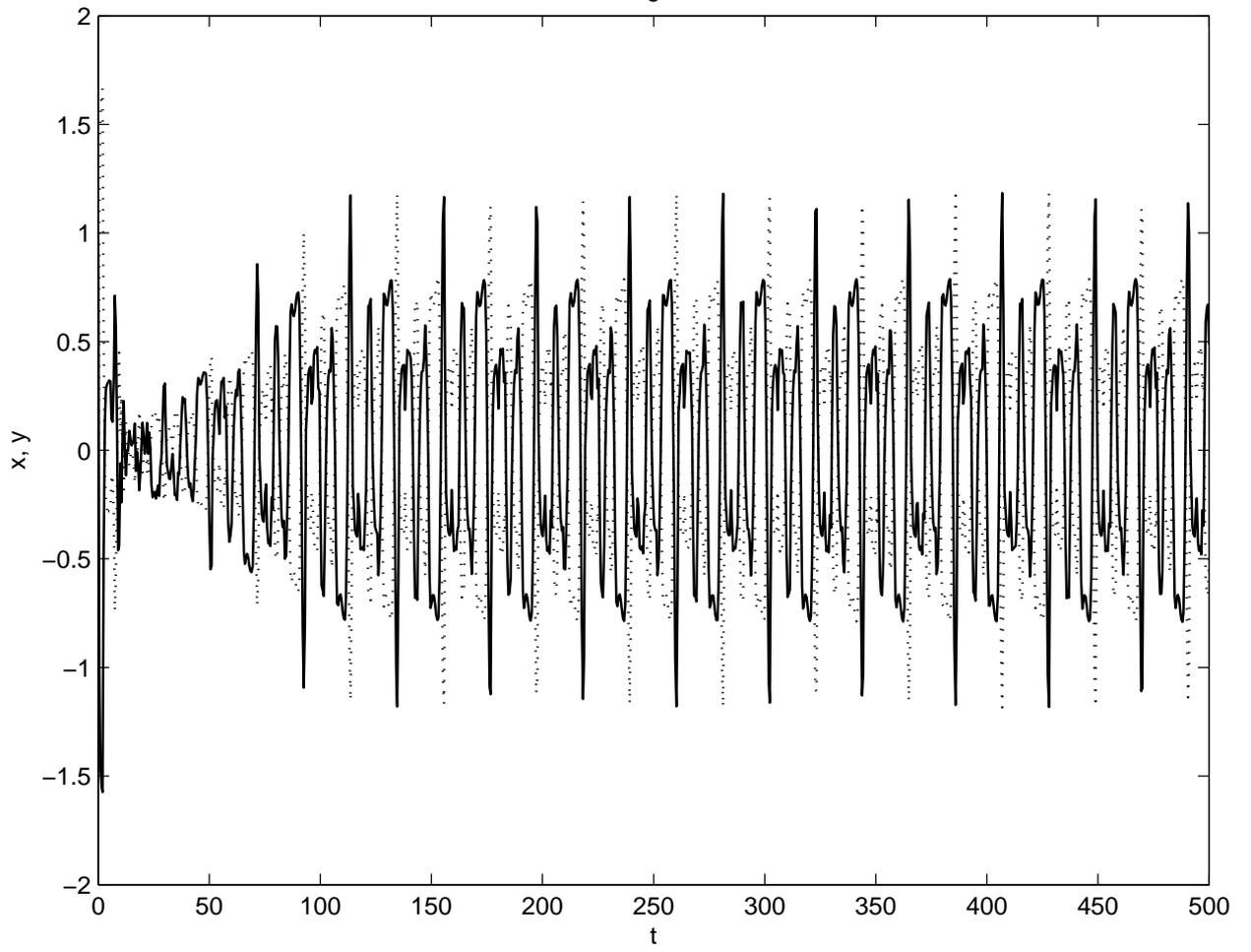

Fig.1

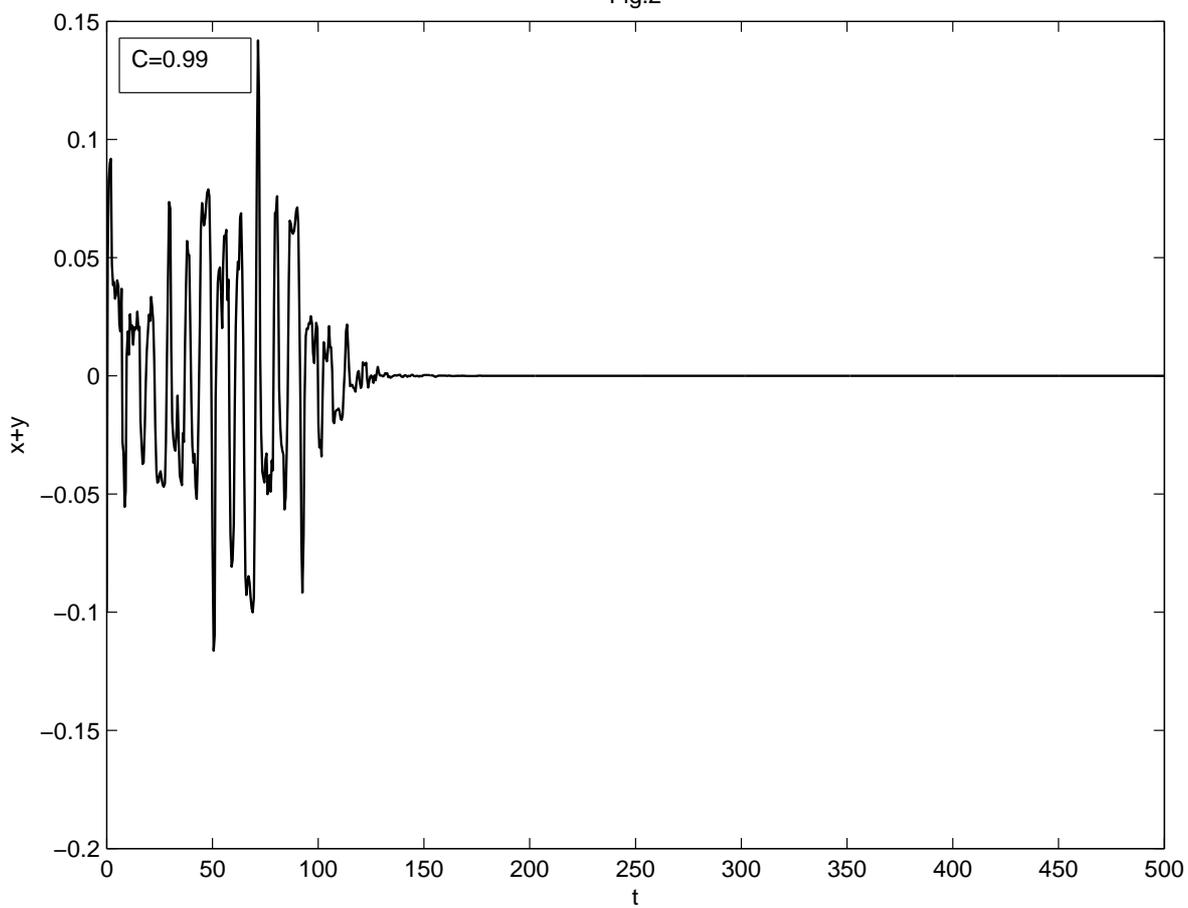

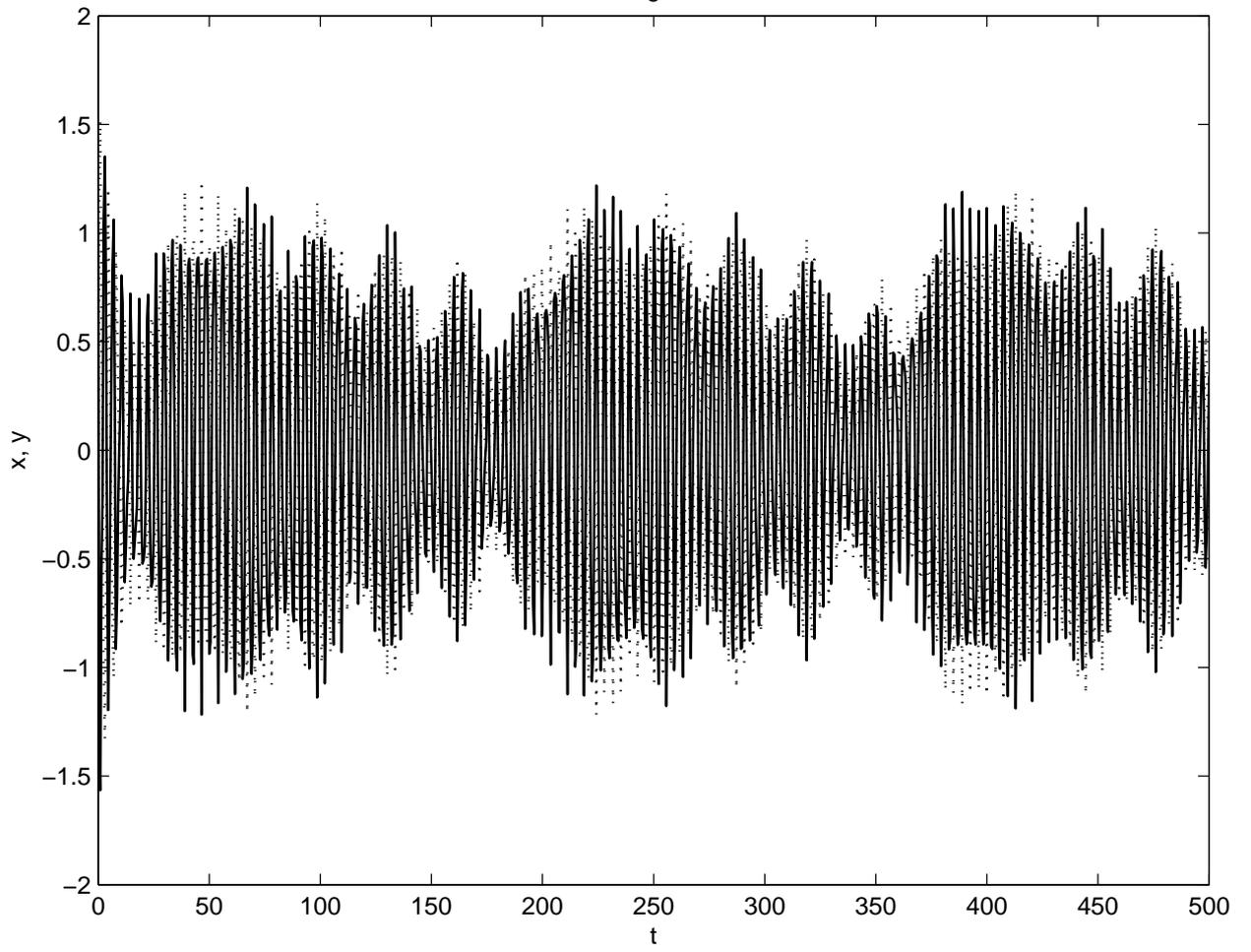

Fig.3

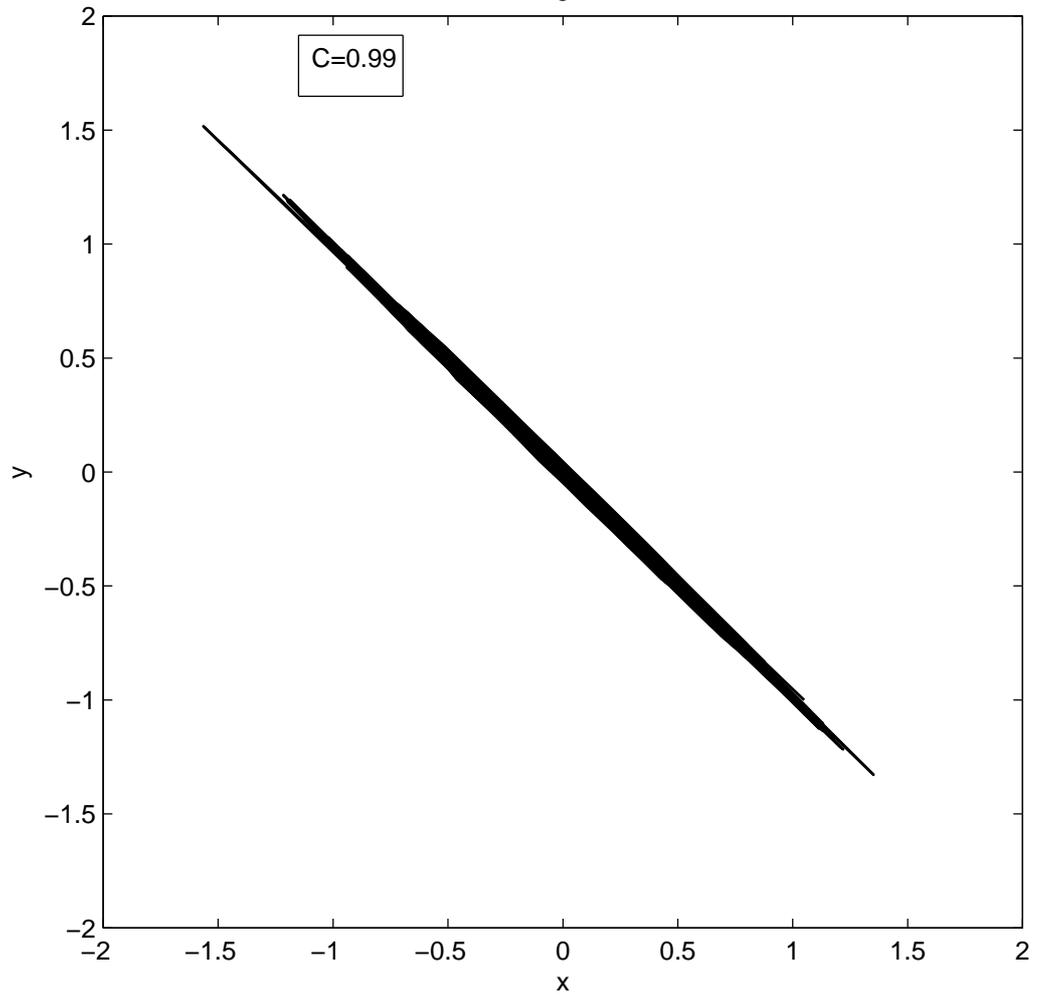

Fig.4